\documentclass[]{elsarticle}

\usepackage{lineno}
\usepackage{amsmath}
\usepackage{graphicx,psfrag,epsf}
\usepackage{enumerate}
\usepackage{natbib}
\usepackage{amsmath,amsfonts,amssymb,graphicx,multirow}
\usepackage{mdsymbol}
\usepackage{booktabs}
\usepackage{amsthm}
\usepackage{bbm}
\newtheorem{remark}{Remark}
\usepackage{algorithm}
\usepackage[noend]{algpseudocode}
\usepackage{rotating}
\modulolinenumbers[5]

\def\spacingset#1{\renewcommand{\baselinestretch}%
{#1}\small\normalsize} \spacingset{1}

\newcommand{\argmax}{\mathop{\mbox{argmax}}}

\newcommand{\comment}[1]{}

\journal{  }

\biboptions{authoryear}

\begin{document}

\begin{frontmatter}

\title{\bf Sparse dimension reduction based on energy and ball statistics}

\author[1]{Emmanuel Jordy Menvouta}
\fntext[1]{KU Leuven, Department of Mathematics, Section of Statistics and Data Science, Leuven, Belgium}

\author[2]{Sven Serneels }
\fntext[2]{Aspen Technology, Bedford, MA, USA} 

\author[1,3]{Tim Verdonck \corref{mycorrespondingauthor}}
\fntext[3]{University of Antwerp, Department of Mathematics, Section of Applied Mathematics, Antwerp, Belgium}
\cortext[mycorrespondingauthor]{Corresponding author}
\ead{Tim.Verdonck@uantwerpen.be}

\begin{abstract}
As its name suggests, sufficient dimension reduction (SDR) targets to estimate a subspace from data that contains all information sufficient to explain a dependent variable. Ample approaches exist to SDR, some of the most recent of which rely on minimal to no model assumptions. These are defined according to an optimization criterion that maximizes a nonparametric measure of association. The original estimators are nonsparse, which means that all variables contribute to the model. However, in many practical applications, an SDR technique may be called for that is sparse and as such, intrinsically performs sufficient variable selection (SVS). This paper examines how such a sparse SDR estimator can be constructed. Three variants are investigated, depending on different measures of association: distance covariance, martingale difference divergence and ball covariance. A simulation study shows that each of these estimators can achieve correct variable selection in highly nonlinear contexts, yet are sensitive to outliers and computationally intensive. The study sheds light on the subtle differences between the methods. Two examples illustrate how these new estimators can be applied in practice, with a slight preference for the option based on martingale difference divergence in the bioinformatics example.  
\end{abstract}


\end{frontmatter}

\section{Introduction}
\label{sec:intro}
Owing to increased data storage, data transmissions speed and computing power, big data analytics have gradually become prevalent over the last five years. The bigger data are, the more difficult it will be to grasp the data by looking at the numbers, or to analyze them visually by plotting fractions of them. That notwithstanding, it can already be a challenge to get a grip on the information contained in small data as well, as soon as dimensionality exceeds what can be visualized. Therefore, dimension reduction has been a mainstay topic of research for the past century and it continues to intrigue researchers up to the present day.

Different approaches to dimension reduction are justified depending on the objectives of the analysis and the structure of the data. Many methods for dimension reduction have been developed to detect structures in one block of data as such, without any external information. A very commonly applied method in that context is principal component analysis (PCA), but sparse variants, or nonlinear methods like autoencoders, will resort under this umbrella as well. However, as soon as there is a dependent variable (or more than one), it becomes more interesting to know which information in the data explains that dependent variable, as opposed to summarizing the data onto a set of latent variables according to a pre-specified criterion, which PCA or autoencoders do. Exactly for this purpose, {\em sufficient dimension reduction} (SDR) has been developed. In SDR, one aims to estimate a set of latent variables that are linear combinations of the original variables $\mathbf{T} = \mathbf{X}\mathbf{W}$ in such a way that the subspace spanned by them contains all information relevant to the dependent variable: 
\begin{equation}
\label{eq:SDR}
\mathbf{Y} \upvDash \mathbf{X}\  | \ \mathbf{T}.
\end{equation}
Here, $\mathbf{X}$ is a sample of $n$ cases of a $p$ variate random variable and $\mathbf{Y}$ is a sample of the dependent variable, whereas $\upvDash$ denotes statistical independence. The space that satisfies \eqref{eq:SDR} is called the {\em central subspace}. There is over twenty years of literature on how the central subspace can be estimated most efficiently according to a wide range of assumptions. The interested reader is referred to \cite{LiSDR} for a recent reference work on SDR. It is not intended to provide a comparison of SDR methods in this article. However, it is noted that some of the most recent approaches to estimate the central subspace are based on maximization of an appropriate nonparametric measure of association. These require no distributional assumptions, and distributional independence \eqref{eq:SDR} follows from absence of association. Because of the latter elegance, that approach will also be the foundational layer for the material developed here, as outlined in Section \ref{sec:SDR}.

Sufficient dimension reduction does identify a subspace whose complement is statistically independent of the predictand. However, SDR is by definition a {\em dense} method: when $\mathbf{X}$ is $p$ variate, so is $\mathbf{W}$ and the elements $w_{ij}$ will generally be nonzero. When all variables can reasonably be assumed to contain information relevant for the predictand $\mathbf{y}$, it makes sense that the information contained in them carries through to the latent variables. However, when {\em uninformative} variables are present, they should be detected as such and they should not contribute to the model, with the corresponding estimated $\hat{w}_{ij}=0 \  \forall i \in [1,p] \  \forall j \in [1,h]$, with $h \in [1,\mathop{\mbox{min}}(n,p)]$ denoting the dimension of the central subspace. Such models are {\em sparse} dimension reduction models and they are in widespread use, with justifications ranging from a better interpretability to the sparse estimate being less prone to noise, just to name a few. Sparse PCA \citep{ZouSPCA2006} is a well established sparse dimension reduction technique. However, sparse sufficient dimension reduction is still a nascent field of research. When uninformative variables are present, the objective becomes to estimate the central subspace in \eqref{eq:SDR} in such a way that the space estimated does not contain contributions from the uninformative variables. Several approaches have been published to achieve variable selection for SDR, either using hypothesis tests based on the estimated latent variables \citep{Cook2004}, called {\em sufficient variable selection} (SVS), or by a sure independence screening \citep{FanLvSIS2008} procedure prior to the analysis. More recently, a sparse method to estimate the central subspace has been proposed, based on a criterion that maximizes distance covariance \citep{SzekelyRizzoBakirov2007} to the dependent variable in the presence of a sparsity penalty \citep{ChenShenYinESSDR2018}. This method relies on fewer assumptions than its predecessors, and as it is based on a nonlinear measure of association, will achieve dimension reduction with respect to nonlinear dependencies. 

This article will build on the sparse SDR method in \cite{ChenShenYinESSDR2018}, but will explore different measures of association. Distance covariance is the oldest association measure that belongs to the class of energy statistics \citep{SzekelyRizzo2013}. However, more recently, further measures of association have been proposed. Martingale difference divergence \citep{ShaoYangMDD2014}, an energy statistic as well, is a related measure of association that, when zero, ascertains conditional mean independence of two stochastic variables. Ball statistics, introduced in the form of sure independence screening in \cite{PanWXZ2019}, on the other hand, comprise an entire different class of statistics, that encompass measures of association as well. While SDR methods based on these more recent measures of association have been developed, sparse variants have not yet been pursued. The latter could be particularly interesting, because the nonsparse variants have been reported to have desirable properties, e.g. SDR based on ball covariance has been reported to be statistically robust and to lead to a straightforward generalization that allows categorical variables as input \citep{ZhangChenBCOVSDR2019}. Based on this motivation, this paper will introduce sparse SDR based on MDD, as well as based on ball covariance. Section will recapitulate the definitions of the association measures and SDR based on them \ref{sec:SDR}. Section \ref{sec:SSDR} will introduce sparse SDR based on these measures and will describe how to compute the resulting estimators. Their properties will be investigated in Section \ref{sec:sim}. Eventually, in Section \ref{sec:Example}, the practical use will be shown for two data set known to contain uninformative variables. The main findings and suggestions for further research are summarized in the conclusion.

\section{Sufficient dimension reduction through energy and ball statistics}
\label{sec:SDR}

As stated in the Introduction, SDR aims to identify a subspace of the original data space by constructing a set of latent variables as linear combinations of the original variables, such that the complement of resulting space spanned by the latent variables is independent of the predictand \eqref{eq:SDR}. There is ample literature on how to estimate the resulting central subspace (see e.g. \cite{LiSDR}). Some of the most recent developments in this direction estimate the central subspace by maximizing a nonparametric and nonlinear measure of association, as that requires minimal assumptions and works in the context of highly nonlinear dependencies. 

To estimate the central subspace using a measure of association $\mathfrak{P}$, the following objective is maximized: 
\begin{subequations}\label{eq:sdrass_empir}
	\begin{equation}\label{eq:sdrass_ctricy}
		\mathbf{W}_h = \argmax_{\mathbf{B}} \mathfrak{P}^2\left(\mathbf{X}\mathbf{B},\mathbf{Y}\right), 
	\end{equation}
	subject to:
	\begin{equation}\label{eq:sdrass_constrcy}
		\mathbf{B}^T\mathbf{X}^T\mathbf{X}\mathbf{B} = \mathbf{I}_h, 
	\end{equation}
\end{subequations}
where $\mathbf{B}$ is an arbitrary $p \times h$ matrix, $h \in [1,\min(n,p)]$. Here, $\mathfrak{P}$ can be any statistic that applies to a set of an $h$ variate and a univariate variable, but of course, in practice, only those statistics estimate a subspace whose complement is independent of $\mathbf{y}$ make sense in this context. As such, three options have been proposed recently: 
\begin{itemize}
    \item SDR by maximizing squared distance covariance \citep{ShengYinDCOVSDR2016};
    \item SDR by maximizing squared martingale difference divergence (MDD, \citeauthor{ZhangMDDSDR2019}, \citeyear{ZhangMDDSDR2019});
    \item SDR by maximizing squared ball covariance \citep{ZhangChenBCOVSDR2019}
\end{itemize}
    
Each of these measures of association will briefly be explained in this Section. 

\begin{remark}
This optimization objective is, in fact, reminiscent of projection pursuit \cite{Huber1985}, targeting to find directions $\mathbf{w}_i$ that maximize a certain objective. In the PP literature, the objective would be called the {\em projection index}. The difference between criterion \eqref{eq:sdrass_ctricy} and the typical PP criterion, is that $\mathfrak{P}$ is allowed to be a statistic that operates on multivariate variables.
\end{remark}
\begin{remark}
Setting $\mathfrak{P}$ to the classical covariance in conjection with a univariate $\mathbf{b}$ will in fact produce the first latent variable of partial least squares (PLS). Successive PLS latent variables can be obtained by maximizing the same objective on deflated data.
\end{remark}

\subsection{Distance covariance}\label{sec:dcov}
Distance covariance (DCOV, \citeauthor{SzekelyRizzo2013}, \citeyear{SzekelyRizzo2013}) is a measure of dependence between random variables that can detect both linear and non linear associations. Let random variables $X \in \mathbb{R}^{p}$ and $Y \in \mathbb{R}$  such that $\mathbb{E}|X|$ and $\mathbb{E}|Y|$ are finite. Here, $p$ is a  positive integer and  the norm $|.|$ is defined   for complex valued functions $f(.), as |f|=f\bar{f}$, where $\bar{f}$ is the complex conjugate of $f$. The squared distance covariance between $X$ and $Y$, is defined as : 

\begin{equation}
    \begin{array}{cc}
    \mathfrak{V}^{2}(X,Y) &= \mathbb{E}|X-X'||Y-Y'|+ \mathbb{E}|X-X'| \mathbb{E}|Y-Y'|\\
                        &-\mathbb{E}|X-X'||Y-Y''|- \mathbb{E}|X-X''||Y-Y'|.
    \end{array}
\end{equation}
   
with $(X',Y')$, $(X'',Y'')$, being iid copies of $(X,Y)$. A fundamental reason for using distance covariance in  \ref{eq:sdrass_ctricy}, is that it has the zero equivalence property in Euclidean spaces : $\mathfrak{V}(X,Y) =0 \iff X \upvDash Y$.   
In practice, given the iid observations $(\mathbf{X},\mathbf{Y})=(\mathbf{X}_{k},\mathbf{Y}_{k})_{k=1}^{n}$,  and the matrix $\mathbf{B} \in \mathbb{R}^{p \times h}$, the sample version of $ \mathfrak{V}^{2}(\mathbf{X} \mathbf{B},\mathbf{Y})$ is given by :
\begin{equation} \label{eq:def_DCOV}
    \mathfrak{V}^{2}(\mathbf{X} \mathbf{B},\mathbf{Y}) = \frac{1}{n^{2}} \sum_{k,l=1}^{n} A_{kl}(\mathbf{B}) C_{kl}.
\end{equation}
In \eqref{eq:def_DCOV}, the entities $A_{kl}$ and $C_{kl}$ are given by:
\begin{subequations}\label{eq:Akldefs_DCOV}
    \begin{equation}
         A_{kl}(\mathbf{B}) = a_{kl}(\mathbf{B})-\bar{a}_{k.}(\mathbf{B})-\bar{a}_{.l}(\mathbf{B}) + \bar{a}_{..}(\mathbf{B}),
    \end{equation}
    where:
    \begin{equation}
         a_{kl}(\mathbf{B}) = |\mathbf{X}_{k}\mathbf{B} - \mathbf{X}_{l} \mathbf{B}|,
    \end{equation}
    \begin{equation}\label{eq:akdot}
        \bar{a}_{k.}(\mathbf{B})= \frac{1}{n} \sum_{l=1}^{n} a_{kl}(\mathbf{B}),
    \end{equation}
    \begin{equation}
        \bar{a}_{.l}(\mathbf{B}) = \frac{1}{n} \sum_{k=1}^{n} a_{kl}(\mathbf{B})
    \end{equation}
    and
    \begin{equation}\label{eq:adotdot}
        \bar{a}_{..}(\mathbf{B}) = \frac{1}{n^{2}}\sum_{k,l=1}^{n} a_{kl}(\mathbf{B}).
    \end{equation}
\end{subequations}

with $|.|$ denoting the Euclidean norm. Similarly, define $c_{kl}=|\mathbf{Y}_{k}-\mathbf{Y}_{l}|$ and $C_{kl} = c_{kl}-\bar{c}_{k.}-\bar{c}_{.l}+\bar{c}_{..}$.

\subsection{Martingale difference divergence}\label{sec:MDD}
Martingale Difference Divergence \citep{ShaoYangMDD2014} is  a measure of conditional mean independence between two random variables. Let $Y \in \mathbb{R}$ and $X \in \mathbb{R}^{p}$ such that $\mathbb{E}(|Y|^{2}+|X|^{2}) < \infty$ and recall that $(X',Y')$ denotes an iid copy of $(X,Y)$. The squared MDD of $Y$ given $X$ is given by  
 \begin{equation}
     \mathfrak{M}^{2}(Y|X) =  - \mathbb{E} \left[(Y - \mathbb{E} Y)(Y' - \mathbb{E}Y')|X - X'|\right].
 \end{equation}

MDD is a particularly useful measure of association between $Y$ and $X$ in presence of heteroskedasticity \citep{ZhangMDDSDR2019}.   Indeed, in that case, maximizing distance covariance will lead to incorrect estimation of the basis for the central subspace. An attractive property of MDD, which makes it particularly useful in SDR, is that $\mathfrak{M}^2(Y|X)=0 \iff \mathbb{E}(Y|X)=\mathbb{E}(Y)$. This is a form of zero equivalence property for the conditional mean. In practice, given the iid observations $(\mathbf{X},\mathbf{Y})=(\mathbf{X}_{k},\mathbf{Y}_{k})_{k=1}^{n}$, and the matrix $\mathbf{B} \in \mathbb{R}^{p \times h}$, the sample version of MDD, $\mathfrak{M}^{2}(Y|X)$,   is given by \citep{park2015}: 
\begin{equation}
    \mathfrak{M}^{2}(\mathbf{Y}|\mathbf{XB}) = \frac{1}{n^{2}} \sum_{k,l=1}^{n} A_{kl}(\mathbf{B}) G_{kl},
\end{equation}
where $A_{kl}$ is defined as in the sample version of DCOV  and 
\begin{subequations}\label{eq:Gkldefs_MDD}
    \begin{equation}
        G_{kl} =g_{kl}-\bar{g}_{k.}-\bar{g}_{.l}+\bar{g}_{..}, 
    \end{equation}
    where:
    \begin{equation}
        g_{kl} = \frac{1}{2}|\mathbf{Y}_{k}-\mathbf{Y}_{l}|^{2}
    \end{equation}
\end{subequations}
and the remaining terms can be found analogously to Equations \eqref{eq:akdot} through \eqref{eq:adotdot}, replacing $a_{kl}$ by $g_{kl}$.

Equations \eqref{eq:Akldefs_DCOV} and \eqref{eq:Gkldefs_MDD}, show that the sample versions of both DCOV and MDD are based on the matrices of pairwise Euclidean distances of $\mathbf{Y}$ and $\mathbf{XB}$. By consequence, the algorithms are $O(n^2)$, with a shortcut available for pairs of univariate variables \citep{HuoFastDCOV2016}. However, in the context of SDR, at least $\mathbf{X}\mathbf{B}$ is typically multivariate and sometimes, so is $\mathbf{Y}$, such that one has to revert to the $O(n^2)$ algorithm.

\subsection{Ball covariance}\label{sec:BCOV}
Ball covariance \citep{PanWXZ2020BCov} is a general measure of dependence between random variables in metric spaces. It was developed  to overcome the fact that distance covariance does not have the zero equivalence property in general metric spaces \citep{Lyons2013}. It has also been shown that distance covariance can be sensible to outliers \citep{RaymaekersRousseeuw2019}, a disadvantage that ball covariance (BCOV) claims not to have. Formally, let
$X$ and $Y$ be two random vectors defined respectively in two separable Banach spaces $(\mathbb{X} , \zeta_{X} )$ and $(\mathbb{Y} , \zeta_{Y})$, where $\zeta_{X}$ and $\zeta_{Y}$
are norms defined on $\mathbb{X}$ and $\mathbb{Y}$ respectively. Let $\theta$, $\mu$ and $\nu$ be Borel probability measures on $\mathbb{X} \times \mathbb{Y}$ , $\mathbb{X}$ and $\mathbb{Y}$ , respectively.
$(X, Y)$ is a Banach-valued random vector defined on a probability space $(\Omega, \mathbb{F}, \mathbb{P})$ such that $(X, Y) \sim \theta$, $X \sim \mu$ and $Y \sim \nu$.
Denote a closed ball with center $x_{1}$ and radius $\zeta_{X}(x_{1},x_{2})$ in $\mathbb{X}$ by $\Bar{B}_{\zeta_{X}}(x_{1},x_{2})$. $\Bar{B}_{\zeta_{Y}}(y_{1},y_{2})$  is then similarly defined in $\mathbb{Y}$.

The squared ball covariance (BCOV) is  defined by 

\begin{equation}
    \mathfrak{B}^{2}(X,Y) = \int \int_{\mathbb{X}\times\mathbb{Y}} [ \theta - \mu \otimes \nu]^{2}(\Bar{B}_{\zeta_{X}}(x_{1},x_{2}) \times \Bar{B}_{\zeta_{Y}}(y_{1},y_{2})) \theta(dx_{1},dx_{2}) \theta(dy_{1},dy_{2}) 
\end{equation}

An interesting property of SDR based on ball covariance is that it is claimed to be robust to outliers \cite{ZhangChenBCOVSDR2019}, even though a formal proof to that claim is not provided (e.g. bounded influence function). Ball covariance also possesses the zero equivalence property : $BCOV(X,Y)=0 \iff X \upvDash Y$, which makes it a usable statistic in the optimization problem for SDR \eqref{eq:sdrass_ctricy}. To compute the sample version of BCOV, let $(\mathbf{X},\mathbf{Y})=(\mathbf{X}_{k},\mathbf{Y}_{k})_{k=1}^{n}$, be an iid sample and given the matrix $\mathbf{B} \in \mathbb{R}^{p \times h}$, then: 
\begin{equation}
    \mathfrak{B}^{2}(\mathbf{X} \mathbf{B},\mathbf{Y}) = \frac{1}{n^6} \sum_{i,j,k,l,s,t=1}^{n} \xi_{ij,klst}^{\mathbf{X} \mathbf{B}} \xi_{ij,klst}^{\mathbf{Y}}.
\end{equation}
Here, 
\begin{subequations}
\begin{equation}
     \xi_{ij,klst}^{\mathbf{X} \mathbf{B}} = \frac{1}{2}(\delta_{ij,kl}^{\mathbf{X} \mathbf{B}} + \delta_{ij,st}^{\mathbf{X} \mathbf{B}} - \delta_{ij,ks}^{\mathbf{X} \mathbf{B}} - \delta_{ij,lt}^{\mathbf{X} \mathbf{B}}),
\end{equation}
\begin{equation}
    \delta_{ij,kl}^{\mathbf{X} \mathbf{B}} = \delta_{ij,k}^{\mathbf{X} \mathbf{B}} \delta_{ij,l}^{\mathbf{X} \mathbf{B}},
\end{equation}
and
\begin{equation}
    \delta_{ij,k}^{\mathbf{X} \mathbf{B}}= \mathbb{I}(\mathbf{X}_{k}\mathbf{B} \in \Bar{B}_{\zeta_{\mathbf{X} \mathbf{B}}}(\mathbf{X}_{i}\mathbf{B}, \mathbf{X}_{j}\mathbf{B}))
\end{equation}
\end{subequations}

and $\xi_{ij,klst}^{\mathbf{Y}}$ is similarly defined. Unlike DCOV and MDD, the computation of the empirical BCOV is not based on the matrix of pairwise Euclidean distances and is more time consuming than either of the other measures of association.  

To obtain sufficient dimension reduction based on DCOV, MDD or BCOV, it suffices to set $\mathfrak{P}$ equal to $\mathfrak{V}$, $\mathfrak{M}$ or $\mathfrak{B}$ in \eqref{eq:sdrass_ctricy}, respectively.

\section{Sparse Sufficient Dimension Reduction}
\label{sec:SSDR}
The SDR methods presented in the previous section suffer from the fact that all the original variables in $\mathbf{X}$ are present in the linear combination $\mathbf{X} \mathbf{B}$. This can be a problem when only a few variables are informative and one would like to select only those variables that are informative. In this section, we consider the extension of the Coordinate Independent Sufficient Variable Selection procedure \citep{chen2010} to the distribution free SDR methods proposed in the previous section. To estimate the central variable selection space, consider maximizing the following objective:

\begin{subequations}\label{eq:svsass_empir}
	\begin{equation}\label{eq:svsass_ctricy}
		\mathbf{W}_h = \argmax_{\mathbf{B}} \mathfrak{P}^2\left(\mathbf{X}\mathbf{B},\mathbf{Y}\right) - \rho(\mathbf{B}), 
	\end{equation}
	subject to:
	\begin{equation}\label{eq:svsass_constrcy}
		\mathbf{B}^T\mathbf{X}^T\mathbf{X}\mathbf{B} = \mathbf{I}_h, 
	\end{equation}
\end{subequations}

where $\rho(\mathbf{B}) = \sum_{i=1}^{p} \theta_{i}|\mathbf{B}_{i}|$, $\mathbf{B}_{i}$ denotes the $i$th row vector of $\mathbf{B}$,  and the $\theta_{i}\geq 0$, $i \in [1,p]$, serve as penalty parameters. Note that in practice, the dimension of the central subspace $h$ needs to be estimated, and to that end the bootstrap method proposed in \cite{ShengYinDCOVSDR2016} can be applied. As statistic  $\mathfrak{P}$, we will consider the three options presented in the previous section, which lead to sufficient variable selection by distance covariance (DCOV---SDR, \citeauthor{ChenShenYinESSDR2018}, \citeyear{ChenShenYinESSDR2018}), Martingale Difference Divergence (MDD---SDR) and Ball Covariance (BCOV---SDR), respectively, the latter two options being introduced in this paper. Note that the sparse sufficient dimension reduction techniques all intrinsically achieve variable selection, and for that purpose, will interchangeably be referred to as {\em sparse variable selection} as well, e.g. DCOV---SVS.

To choose the penalty parameters, the approach from \cite{ChenShenYinESSDR2018} is followed. Consider parameters of the form
\begin{equation}
    \theta_{i} = \theta|\mathbf{B}_{i}|^{-a},
\end{equation}
where $\mathbf{B}_{i}$ is the $i$th row vector of the solution to problem \eqref{eq:sdrass_empir} and $a$ is a real number in $[0, 1]$. This strategy is practical, since it turns a $p$-dimensional tuning parameter problem into a one-dimensional problem. To choose the tuning parameter $\theta$, a BIC-type criterion is maximized \citep{LiSVS2007}:
\begin{equation}
     \argmax_{\theta} \log\left(\mathfrak{P}^2\left(\mathbf{X}\tilde{\mathbf{B}}_{\theta},\mathbf{Y}\right)\right) - \frac{\log(n)(p_{\theta}-h)h}{n},
     \label{eq:thetasol}
\end{equation}

with $\tilde{\mathbf{B}}_{\theta}$ denoting the sparse solution given by $\theta$ and $p_{\theta}$ the number of nonzero rows of $\tilde{\mathbf{B}}_{\theta}$. Given that $\rho$ is not differentiable, the local quadratic approximation (LQA) is applied as in \cite{FanLi2001} and the penalty function is approximated by a quadratic function at every step of the iteration. More precisely, if $\mathbf{B}^{n}=(\mathbf{B}_{1}^{n}, \ldots, \mathbf{B}_{p}^{n})^{T}$ is the solution to of the optimization problem in \eqref{eq:svsass_empir} at the $n$th  iteration, then the first derivative of $\rho(\mathbf{B})$ around $\mathbf{B}^{n}$ can be approximated as:
\begin{equation}
    \frac{\partial \rho}{\partial \mathbf{B}} \approx \mathop{\mbox{diag}}\left(\frac{\theta|\mathbf{B}^{n}_{1}|^{-a}}{|\mathbf{B}^{n}_{1}|}, \ldots, \frac{\theta|\mathbf{B}^{n}_{p}|^{-a}}{|\mathbf{B}^{n}_{p}|} \right) \mathbf{B} := \mathbf{H}^{n} \mathbf{B}.
\end{equation}
Using a second-order Taylor expansion  of $\rho(\mathbf{B})$ about $\mathbf{B}^{n}$ results in  
\begin{equation}
 \rho(\mathbf{B}) \approx \frac{1}{2} \mathop{\mbox{tr}}(\mathbf{B}^{T} \mathbf{H}^{n} \mathbf{B}) + C_{n},
\end{equation}
with $C_{n}$ a constant with respect to $\mathbf{B}$. This LQA is then plugged in \eqref{eq:svsass_ctricy} to obtain the sparse SDR procedure presented in Algorithm \ref{sparseSDRalgo} below. In practice, $a$ is set to 0.5, 0.2 and 0.8 for DCOV---SVS, MDD---SVS and BCOV---SVS, respectively. It is practicable to fix the upper bound of iterations $N$ to 200. The tolerance $\tau$ is set 0.001 as in \cite{ChenShenYinESSDR2018} and the starting point $\mathbf{B}^{0}$ is taken to be the nonsparse solution from \eqref{eq:sdrass_empir} obtained in python through the direpack package \citep{menvouta2020}.To choose the penalty parameter $\theta$,  values in $[0,0.5]$ spaced by $0.01$ are considered in algorithm \ref{sparseSDRalgo} and the solution to \eqref{eq:thetasol} is selected. When the  structural dimension $h$ is unknown, it can be estimated using the bootstrap method of \cite{ShengYinDCOVSDR2016} with $200$ replications. The \texttt{subspaceAngle} method in algorithm \ref{sparseSDRalgo} refers to the principal angle between subspaces of \cite{Knyazev2002PrincipalAB}.

\begin{algorithm}[!htbp]
\caption{SVS algorithm}
\begin{algorithmic}[2]

\Procedure{sparseSDR}{$\mathbf{X},\mathbf{Y}, \mathbf{B}^{0},h,\theta, \tau, a,N$}     
    \State  $\mathbf{s}  \leftarrow \mathbf{0}_{p}$.  
    \State $\mathbf{V}^{0} \leftarrow \mathbf{B}^{0}$
    \State $\mathbf{X}^{0} \leftarrow \mathbf{X}$
    \For{$i = 1$ to $N$}
    \State $\mathbf{H}^{0} \leftarrow \mathop{\mbox{diag}}\left(\frac{\theta|\mathbf{V}^{0}_{1}|^{-a}}{|\mathbf{V}^{0}_{1}|}, \ldots, \frac{\theta|\mathbf{V}^{0}_{p}|^{-a}}{|\mathbf{V}^{0}_{p}|} \right)$
     \State $ \mathbf{V} \leftarrow  \mathop{\mbox{argmax}}_{\mathbf{V}} \mathfrak{P}^2\left(\mathbf{X}^{0}\mathbf{V},\mathbf{Y}\right) - \frac{1}{2} \mathop{\mbox{tr}}(\mathbf{V}^{T} \mathbf{H}^{0} \mathbf{V})$ s.t. $\mathbf{V}^T\mathbf{X}^{0,T}\mathbf{X}^{0}\mathbf{V} = \mathbf{I}_h$
     \If{subspaceAngle$(\mathbf{V}^{0},\mathbf{V}) \leq \tau$}  break
     \EndIf
    \For{$j=1$ to $p$}
        \If{$|\mathbf{V}_{j}| \leq \tau$}
        \State $\mathbf{s}_{j} \leftarrow 1$
        \EndIf
    \EndFor
    \If{ $\mathbf{s}=\mathbf{1}_{p}$}  break
     \EndIf
     \State $l \leftarrow 1$
     \For{$j=1$ to $p$}
        \If{$\mathbf{s}_{j} = 0$}
        \State   $\mathbf{V}^{0}_{l} \leftarrow \mathbf{V}_{j}$
        \State  $\mathbf{X}^{0}_{l} \leftarrow \mathbf{X}_{j}$
        \State $l \leftarrow l+1$
        \EndIf
    \EndFor
    \EndFor
    
    \State $\mathbf{V}^{sol} \leftarrow \mathbf{0} \in \mathbb{R}^{p \times h}$
    \State $k \leftarrow 0$
    \For{$j=1$ to $p$}
        \If{$\mathbf{s}_{j}=0$}
        \State $\mathbf{V}^{sol}_{j} \leftarrow \mathbf{V}^{0}_{k}$
        \State $k \leftarrow k+1$
        \EndIf
    \EndFor
    
    \State return $\mathbf{V}^{sol}$
   
\EndProcedure

\end{algorithmic}
\label{sparseSDRalgo}
\end{algorithm}

\section{Simulation study}
\label{sec:sim}

\subsection{Simulation and evaluation setup}

This section aims to answer a few questions regarding the properties of different SVS methods in a data driven way. The main question to be investigated is up to which extent the individual SVS methods are able to identify the right subset of informative variables, based on data generated according to different mechanisms. As such, the methods from Section \ref{sec:SSDR} will be compared as well to established sparse variable selection tools. For the latter purpose, Sparse NIPALS, also called SNIPLS  \citep{SPRM-DA}, as well as its outlier robust counterpart, sparse partial robust M (SPRM, \citeauthor{SPRM}, \citeyear{SPRM}), are selected. SPRM is a sparse variant of the older partial robust M-regression \citep{PRM}, a well established robust alternative to partial least squares, which on its turn is proven to estimate the central subspace \citep{CookHellandSu2013}. This will allow to evaluate up to which extent the different methods perform well when the data are generated according to linear and nonlinear models, as well as with or without outliers. 

\comment{
Based on theoretical considerations and previous literature results, it can be expected that all methods perform well when data are generated according to a linear model and no outliers are present. SPRM should still perform well for data generated by a linear model in the presence of outliers. However, when data are generated from a nonlinear model, which will be the majority of the cases considered, one can expect that both SNIPLS and SPRM will perform more poorly, while all three SVS methods described here still perform well. Finally, since \cite{ZhangChenBCOVSDR2019} claim that BCOV-SDR is outlier robust. This section will investigate up to which extent that carries through to the SVS context, and up to which extent the DCOV and MDD based variants may or may not be robust.}

The optimization problem \eqref{eq:svsass_empir} is solved by dint of IPOPT \citep{WaechterBiegler2006}, with the MUMPS solver embedded. Setting $n=120$ $p=24$,  assuming that the dimension $h$ of the central subspace is known, and considering standard normally distributed error terms independent from the predictors, eight different data models are investigated by generating a hundred simulated datasets from each data model. The performance of each of the proposed methods is investigated for the different data models. To assess the quality of the methods, three metrics are reported: 
\begin{itemize}
    \item the True Positive Rate (TPR), defined as the proportion of correctly identified active variables,
    \item the False Positive Rate (FPR), defined as the proportion of inactive variable identified as active and 
    \item the $F_1$-score,  $F_1= \frac{\mathop{\mbox{tp}}}{\mathop{\mbox{tp}} +0.5(\mathop{\mbox{fp}}+\mathop{\mbox{fn}})}$, where $\mathop{\mbox{tp}}$ denotes the number of correctly identified active variables, $\mathop{\mbox{fp}}$ the number of inactive variables identified as active and $\mathop{\mbox{fn}}$ the number of active variables identified as inactive. 
\end{itemize}
Beyond reporting the $F_1$-score for each method, the Kruskal-Wallis non parametric ANOVA \citep{KruskalWallis1952} and the Conover \citep{Conover1999} post-hoc tests are performed. This allows to take into account the uncertainty and error in the estimates when comparing different methods. Methods perform similarly if the $p$-value of the corresponding pairwise post-hoc test is greater than 0.05.

\newpage
\subsection{Data generating models}

\begin{center}
\begin{sidewaystable}[!ht]
\begin{tabular}{cccc}
\hline\hline
Study & \multicolumn{1}{c}{\textbf{X}}  & \multicolumn{1}{c}{\textbf{$\mathbf{B}^{T}$}}& \multicolumn{1}{c}{\textbf{Y}} \\
\hline
\multirow{2}{*}{A} & $\mathbf{X} \sim N(\mathbf{0}_{p}, \boldsymbol{\Sigma})$ & $\mathbf{B}_{1}^{T} =(1,0,0, \ldots, 0)$ &$ \frac{\mathbf{X} \mathbf{B}_{1}}{0.5 +(\mathbf{X} \mathbf{B}_{2} + 1.5)^2} +0.2 \boldsymbol{\epsilon}$ \\
&  &$\mathbf{B}_{2}^{T}=(0,1,0, \ldots, 0)$  & \\
\hline 
\multirow{2}{*}{B} &  $(\mathbf{x}_{2},\ldots,\mathbf{x}_{p}) \sim N(\mathbf{0}_{p-1}, \boldsymbol{\Sigma})$  &$\mathbf{B}_{1}^{T} =0.5(1,1,1,1,0,\ldots,0)$ & $ (\mathbf{X} \mathbf{B}_{1})^{2} + \mathbf{X} \mathbf{B}_{2} + 0.5\boldsymbol{\epsilon}_{1} $  \\
&  $\mathbf{x}_{1}=|\mathbf{x}_{2}+\mathbf{x}_{3}| +\boldsymbol{\epsilon}_{2}$ &$\mathbf{B}_{2}^{T} =0.5(1,-1,1,-1,0,\ldots,0)$ &\\
\hline 
\multirow{2}{*}{C}& $\mathbf{X} \sim N(\mathbf{0}_{p}, \boldsymbol{\Sigma})$ &$\mathbf{B}_{1}^{T} = (1,1,1,1,0,\ldots,0)$ & $ \mathop{\mbox{sign}}( 2 \mathbf{X} \mathbf{B}_{1} + \boldsymbol{\epsilon}_{1}) \times$\\
&  &$\mathbf{B}_{2}^{T} = (1,-1,1,-1,0,\ldots,0)$&$ \log| 2 \mathbf{X} \mathbf{B}_{2} + 4+\boldsymbol{\epsilon}_{2}|$\\
\hline 
\multirow{2}{*}{D}& $\mathbf{X} \sim N(\mathbf{0}_{p}, \boldsymbol{\Sigma})$ &$\mathbf{B}_{1}^{T} = (1,1,1,1,0,\ldots,0)$ & $\mathbb{I}(\mathbf{X} \mathbf{B}_{1}+0.2 \boldsymbol{\epsilon} >1)$  + \\
&  &$\mathbf{B}_{2}^{T} = (0,0,0,0,1,1,1,1,0\ldots,0)$ &  $2 \mathbb{I}(\mathbf{X} \mathbf{B}_{2} +0.2 \boldsymbol{\epsilon} >0)$\\
\hline 
\multirow{2}{*}{E}& $\mathbf{X} \sim N(\mathbf{0}_{p}, \boldsymbol{\Sigma})$  &$\mathbf{B}_{1}^{T} = (1,1,1,1,0,\ldots,0)$ & $(\mathbf{X} \mathbf{B}_{1})^{2} + |\mathbf{X} \mathbf{B}_{2}| +$\\
&  &$\mathbf{B}_{2}^{T} = (1,-1,1,-1,0,\ldots,0)$ & $0.1\boldsymbol{\epsilon}$\\
\hline
\multirow{2}{*}{F}& $\mathbf{X} \sim N(\mathbf{0}_{p}, \boldsymbol{\Sigma})$  &$\mathbf{B}_{1}^{T} = (1,1,1,1,0,\ldots,0)$ & $(\mathbf{X} \mathbf{B}_{1})^{2} + |\mathbf{X} \mathbf{B}_{2}| +0.1\boldsymbol{\epsilon}_{1}$\\
& ($\mathbf{x}_1$, $\mathbf{x}_2$, $\mathbf{x}_3$, $\mathbf{x}_4$) $\sim 2 \cdot t_{1}$   &$\mathbf{B}_{2}^{T} = (1,-1,1,-1,0,\ldots,0)$ & \\
\hline
\multirow{2}{*}{G}& $\mathbf{X} \sim N(\mathbf{0}_{p}, \boldsymbol{\Sigma})$  &$\mathbf{B}_{1}^{T} = (1,1,1,1,0,\ldots,0)$ & $(\mathbf{X} \mathbf{B}_{1})^{2} + |\mathbf{X} \mathbf{B}_{2}| +0.1\boldsymbol{\epsilon}_{1}$\\
& ($\mathbf{X}_5$, $\mathbf{X}_6$,$\mathbf{X}_7$,$\mathbf{X}_8$) $\sim 2 \cdot t_{1}$   &$\mathbf{B}_{2}^{T} = (1,-1,1,-1,0,\ldots,0)$ & \\
\hline
\multirow{2}{*}{H}& $\mathbf{X} \sim N(\mathbf{0}_{p}, \boldsymbol{\Sigma})$  &$\mathbf{B}_{1}^{T} = (1,1,1,1,0,\ldots,0)$ & $(\mathbf{X} \mathbf{B}_{1})^{2} + |\mathbf{X} \mathbf{B}_{2}| +0.1\boldsymbol{\boldsymbol{\epsilon}}_{1}$\\
& ($\mathbf{x}_9$, $\mathbf{x}_{10}$, $\mathbf{x}_{11}$, $\mathbf{x}_{12}$) $\sim 2 \cdot t_{1}$   &$\mathbf{B}_{2}^{T} = (1,-1,1,-1,0,\ldots,0)$ &\\
  \hline\hline
\end{tabular}
\caption{Simulation studies setup. \label{table:simstudies} }
\end{sidewaystable}
\end{center}

Table \ref{table:simstudies} shows the distribution of $\mathbf{X}$, the structure of the covariance matrix $\boldsymbol{\boldsymbol{\Sigma}}$, the basis for the central subspace $\mathbf{B}$ and the expression for $\mathbf{Y}$ for each of the eight simulation study setups. The error terms $\boldsymbol{\boldsymbol{\epsilon}}$, $\boldsymbol{\boldsymbol{\epsilon}}_{1}$ and $\boldsymbol{\boldsymbol{\epsilon}}_{2}$ follow independent standard normal distributions and are independent from $\mathbf{X}$. For studies A, B and E we have $\boldsymbol{\Sigma}_{i,j} = 0.5^{|i-j|}$ and for studies C and D $\boldsymbol{\Sigma}$ is taken to be the identity matrix. The setup in Study A was taken from \cite{ChenShenYinESSDR2018}, in study B from \cite{ZhangMDDSDR2019}, in Study C from \cite{ShengYinDCOVSDR2016}, in Study D from \cite{YinHilafu2015} and in Study E from \cite{ZhangChenBCOVSDR2019}. Studies F, G and H are set up as Study E, but 10\% of some variables are replaced by outliers $2\cdot t_{1}$, where $t_{1}$ is a Cauchy distributed random variable. The aim is to investigate which method performs best when outliers are present in the informative and/or uninformative variables, or both. In study F, outliers are added in the informative variables alone, which presumably should still lead to those variables being detected as informative and contributing to the model. Study G investigates the effect of outlying cells in the uninformative variables, which should not be selected as informative. Study H generates casewise outliers in both informative and uninformative variables.

\subsection{Results}

The simulation results for the eight studies are presented in Tables \ref{table:svsf1} through \ref{table:svsfpr}. One cannot expect a specific method to have better performance than all the other ones across all simulation studies. However, the  SVS methods obtained from solving (\ref{eq:svsass_empir}) are superior to the linear dimension reduction methods {SNIPLS} and SPRM as they return a better $F_1$-score than the linear methods across all 8 studies.

Studies A, B and C illustrate that increasing the level of non-linearity in the expression of $\mathbf{Y}$ favors DCOV---SVS and BCOV---SVS over MDD---SVS. Effectiveness with respect to nonlinearity is reflected in their $F_{1}$-scores and FPRs, which are consistently better than those of MDD---SVS. The fact that MDD---SVS performs worst is to be expected as DCOV and BCOV can detect more general forms of associations than MDD. Accounting for the mean and standard deviation of the estimated $F_{1}$-score, DCOV-SVS and BCOV-SVS perform similarly. 

In the sparse categorical setting (Study D) DCOV---SVS, MDD---SVS and BCOV---SVS perform similarly in terms of $F_{1}$-score. This is an indication that although these non-linear methods are more general and are able to detect general forms of associations, they perform better when the dependent variable is continuous. 

Studies E through H assess the effect outliers have on the resulting sufficient variable selection. In the absence of outliers (Study E), DCOV-SVS performs best as it is able to correctly detect the nonlinear associations between $\mathbf{X}$ and $\mathbf{Y}$.  Study F presents the case where outliers are added to the active variables. In this case, DCOV---SVS and BCOV---SVS perform similarly with an average $F_1$-score of $0.921$ and $0.923$, respectively. This implies that both methods show similar sensitivity to outliers in the active variables. When outliers are added to the inactive variables (Study G), DCOV---SVS performs best with an $F_{1}$-score of $0.896$, showing that it is less sensitive to outliers in the inactive variables than both other variants. When outliers are added to all variables (Study H), the 3 nonlinear methods and SNIPLS perform similarly.  This indicates that all the nonlinear methods under consideration are the most sensible to outliers when these outliers occur on a casewise basis.  

Tables \ref{table:svstheta} and \ref{table:svstime} show the average penalty parameter and the average time taken for one simulation run in seconds.  Regarding the penalty parameter $\theta$, on average BCOV---SVS selects a lower penalty parameter than both other options. Howbeit, in 6 of the 8 studies considered,  BCOV---SVS takes longer to fit. That said, when outliers are present in the active variables, MDD---SVS takes longer to fit. Expectedly, SPRM and SNIPLS are much faster to fit than the other methods, since they do not rely on a numerical nonlinear optimization.

\begin{table}
\centering
\begin{tabular}{ ccccccccc}
\hline\hline
 & \multicolumn{2}{c}{\textbf{A}} & \multicolumn{2}{c}{\textbf{B}} & \multicolumn{2}{c}{\textbf{C}} & \multicolumn{2}{c}{\textbf{D}}\\
 \cmidrule(lr{1em}){2-3} \cmidrule(lr{1em}){4-5} \cmidrule(lr{1em}){6-7} \cmidrule(lr{1em}){8-9} 
   Method& mean & sd & mean & sd & mean & sd & mean & sd\\
 \midrule
  {DCOV-SVS}  &0.910& 0.178  &0.782 &0.126 & 0.824 & 0.124 & 0.698 & 0.083  \\
  MDD-SVS&0.824 &0.223 &\textbf{0.863} &0.150 & 0.794 & 0.157 & \textbf{0.706} & 0.089 \\
  BCOV-SVS &\textbf{0.925} &0.165 &0.792 &0.228 & \textbf{0.848} & 0.174 & 0.680 & 0.226 \\
  SNIPLS &0.264 &0.109 &0.720 &0.213 & 0.557 &0.190 & 0.607 & 0.117\\
  SPRM &0.281 &0.140 &0.592 &0.141& 0.552 & 0.185 &0.585 & 0.140\\
  
  \hline 
  
& \multicolumn{2}{c}{\textbf{E}} & \multicolumn{2}{c}{\textbf{F}} & \multicolumn{2}{c}{\textbf{G}} & \multicolumn{2}{c}{\textbf{H}}\\
 \cmidrule(lr{1em}){2-3} \cmidrule(lr{1em}){4-5} \cmidrule(lr{1em}){6-7} \cmidrule(lr{1em}){8-9}  
   Method& mean & sd & mean & sd & mean & sd & mean & sd\\
 \midrule
  {DCOV-SVS}  &\textbf{0.965} &0.091   &0.921  &0.151& \textbf{0.896} &0.186 &0.262 &0.009  \\
  MDD-SVS &0.699 &0.160 &0.468 & 0.140 &0.700 & 0.155 & \textbf{0.290} &0.007 \\
  BCOV-SVS &0.689 &0.200 & \textbf{0.923} & 0.082 &0.597 & 0.227& 0.283 &0.002 \\
  SNIPLS &0.282 &0.159 &0.637 & 0.204 &0.279 &0.164 &0.286 & 0.000\\
  SPRM &0.155 &0.162 &0.161 & 0.148 &0.180 &0.168 &0.210 &0.017\\
  
  \hline\hline
\end{tabular}
\caption{Mean and standard deviation of F1 score. \label{table:svsf1} }
\end{table}

\begin{table}
\centering
\begin{tabular}{ ccccccccc}
\hline\hline
 & \multicolumn{2}{c}{\textbf{A}} & \multicolumn{2}{c}{\textbf{B}} & \multicolumn{2}{c}{\textbf{C}} & \multicolumn{2}{c}{\textbf{D}}\\
 \cmidrule(lr{1em}){2-3} \cmidrule(lr{1em}){4-5} \cmidrule(lr{1em}){6-7} \cmidrule(lr{1em}){8-9} 
   Method& mean & sd & mean & sd & mean & sd & mean & sd\\
 \midrule
  {DCOV-SVS}  &\textbf{0.910}&0.193&0.688 &0.172 & 0.752 & 0.196 & 0.550 & 0.114 \\
  MDD-SVS&0.825 &0.240 &\textbf{0.853} &0.138 & 0.762 & 0.196 & 0.566 & 0.126  \\
  BCOV-SVS &0.895 &0.217 &0.747 &0.276 & 0.820 & 0.256 & 0.595 &0.284  \\
  SNIPLS &0.668 &0.287 &0.795 &0.193 &\textbf{0.885} & 0.205 & \textbf{0.940} & 0.159  \\
  SPRM &0.513 &0.298 &0.625 &0.199 & 0.883 & 0.190 & 0.910 & 0.220  \\
  
  \hline 
  
& \multicolumn{2}{c}{\textbf{E}} & \multicolumn{2}{c}{\textbf{F}} & \multicolumn{2}{c}{\textbf{G}} & \multicolumn{2}{c}{\textbf{H}}\\
 \cmidrule(lr{1em}){2-3} \cmidrule(lr{1em}){4-5} \cmidrule(lr{1em}){6-7} \cmidrule(lr{1em}){8-9}  
   Method& mean & sd & mean & sd & mean & sd & mean & sd\\
 \midrule
  {DCOV-SVS}  &0.998 &0.023 &0.920 &0.122 & 0.990 & 0.061 & 0.578 &0.240    \\
  MDD-SVS &\textbf{1.000} & 0.000 &0.970 &0.082 & \textbf{0.998} & 0.025& 0.760 & 0.181  \\
  BCOV-SVS &0.760 &0.293 &0.878 &0.135 & 0.8275 & 0.320 &0.978 & 0.072  \\
  SNIPLS &0.530 &0.365 &\textbf{1.000} &0.000 &0.525 & 0.353 &\textbf{1.000} & 0.000  \\
  SPRM &0.253 &0.292 &0.320 &0.348 &0.285 & 0.297 &0.388 &0.363 \\
  
  \hline\hline
\end{tabular}
\caption{Mean and standard deviation of TPR. \label{table:svstpr} }
\end{table}

\begin{table}
\centering
\begin{tabular}{ ccccccccc}
\hline\hline
 & \multicolumn{2}{c}{\textbf{A}} & \multicolumn{2}{c}{\textbf{B}} & \multicolumn{2}{c}{\textbf{C}} & \multicolumn{2}{c}{\textbf{D}}\\
 \cmidrule(lr{1em}){2-3} \cmidrule(lr{1em}){4-5} \cmidrule(lr{1em}){6-7} \cmidrule(lr{1em}){8-9} 
   Method& mean & sd & mean & sd & mean & sd & mean & sd\\
 \midrule
  {DCOV-SVS}  &0.023&0.141&\textbf{0.017} &0.101 & \textbf{0.010} & 0.029 & \textbf{0.007} & 0.026  \\
  MDD-SVS&0.033 &0.143 &0.051 &0.197 & 0.050 & 0.176 & 0.016 & 0.101 \\
  BCOV-SVS &\textbf{0.001} &0.01 &\textbf{0.017} &0.038 & 0.013 & 0.025 & 0.032 & 0.059 \\
  SNIPLS &0.395 &0.322 &0.167 &0.322 & 0.357 & 0.324 & 0.597 & 0.262\\
  SPRM &0.262 &0.284 &0.154 &0.268  & 0.369 &0.336 & 0.598 & 0.281\\
  
  \hline 
  
& \multicolumn{2}{c}{\textbf{E}} & \multicolumn{2}{c}{\textbf{F}} & \multicolumn{2}{c}{\textbf{G}} & \multicolumn{2}{c}{\textbf{H}}\\
 \cmidrule(lr{1em}){2-3} \cmidrule(lr{1em}){4-5} \cmidrule(lr{1em}){6-7} \cmidrule(lr{1em}){8-9}  
   Method& mean & sd & mean & sd & mean & sd & mean & sd\\
 \midrule
  {DCOV-SVS}  &\textbf{0.023} &0.102 &0.042 &0.197 & \textbf{0.093} & 0.252 &0.557 &0.160  \\
  MDD-SVS &0.219 &0.223 &0.483 &0.188 &0.212 & 0.205 & 0.699&0.010 \\
  BCOV-SVS &0.087 &0.114 &\textbf{0.003} &0.017 &0.190 &0.155 &0.987 &0.031  \\
  SNIPLS &0.419 &0.300 &0.313 & 0.279 &0.420 & 0.286 & 1.000 &1.000 \\
  SPRM &0.322 &0.277 &0.404 &0.307 & 0.307 & 0.248 & \textbf{0.389} &0.332 \\
  
  \hline\hline
\end{tabular}
\caption{Mean and standard deviation of FPR. \label{table:svsfpr} }
\end{table}

\begin{table}
\centering
\begin{tabular}{ ccccccccc}
\hline\hline
 & \multicolumn{2}{c}{\textbf{A}} & \multicolumn{2}{c}{\textbf{B}} & \multicolumn{2}{c}{\textbf{C}} & \multicolumn{2}{c}{\textbf{D}}\\
 \cmidrule(lr{1em}){2-3} \cmidrule(lr{1em}){4-5} \cmidrule(lr{1em}){6-7} \cmidrule(lr{1em}){8-9} 
   Method& mean & sd & mean & sd & mean & sd & mean & sd\\
 \midrule
  {DCOV-SVS}  &0.234& 0.167  &0.169  &0.093 & 0.180 &0.100 & 0.134 & 0.053 \\
  MDD-SVS&0.211 &0.156 &0.328 &0.116 & 0.260 & 0.102 &0.221 & 0.008 \\
  BCOV-SVS &\textbf{0.136} &0.152 &\textbf{0.063} &\textbf{0.091} & \textbf{0.131} & 0.147 &\textbf{0.061} &0.100 \\
  SNIPLS &0.486 &0.291 &0.557 &0.274 & 0.527 & 0.286 & 0.305 & 0.158\\
  SPRM &0.600 &0.280 &0.553 &0.247 & 0.519 & 0.293 & 0.297 & 0.180\\
  
  \hline 
  
& \multicolumn{2}{c}{\textbf{E}} & \multicolumn{2}{c}{\textbf{F}} & \multicolumn{2}{c}{\textbf{G}} & \multicolumn{2}{c}{\textbf{H}}\\
 \cmidrule(lr{1em}){2-3} \cmidrule(lr{1em}){4-5} \cmidrule(lr{1em}){6-7} \cmidrule(lr{1em}){8-9}  
   Method& mean & sd & mean & sd & mean & sd & mean & sd\\
 \midrule
  {DCOV-SVS}  &0.177 &0.074   &0.345 &0.128 &0.174 &0.101 &0.388 &0.139  \\
  MDD-SVS &0.413 &0.120 &0.264 &0.155 &0.415 &0.116 & 0.219 & 0.132\\
  BCOV-SVS & \textbf{0.054} &0.079 &\textbf{0.120} &0.125 &\textbf{0.042} &0.076 &\textbf{0.201} &0.166 \\
  SNIPLS &0.604 &0.261 &0.452 &0.277 &0.514 &0.295 &0.526 &0.269\\
  SPRM &0.681 &0.260 &0.601 &0.257 &0.670 &0.220 &0.604 &0.279\\
  
  \hline\hline
\end{tabular}
\caption{Mean and standard deviation of the sparsity parameter. \label{table:svstheta} }
\end{table}

\begin{sidewaystable}
\centering
\begin{tabular}{ccccccccc}
\hline\hline
 & \multicolumn{2}{c}{\textbf{A}} & \multicolumn{2}{c}{\textbf{B}} & \multicolumn{2}{c}{\textbf{C}} & \multicolumn{2}{c}{\textbf{D}}\\
 \cmidrule(lr{1em}){2-3} \cmidrule(lr{1em}){4-5} \cmidrule(lr{1em}){6-7} \cmidrule(lr{1em}){8-9} 
   Method& mean & sd & mean & sd & mean & sd & mean & sd\\
 \midrule
  {DCOV-SVS}  &6645.4& 4118.2  &7935.7 &5441.6 &9890.6 & 5129.0 & 8253.7 & 6228.4  \\
  MDD-SVS&9048.5 &5183.0 &9442.9 &6801.1 & 12470.9 & 7092.2 &10584.7 & 8023.9 \\
  BCOV-SVS &\textbf{10068.3} &3303.5 &\textbf{13089.6} &3343.7 & \textbf{14138.0} & 5837.5 & \textbf{22440.7} & 7336.7 \\
  SNIPLS &0.1 &0.1 &0.01 &0.0 & 0.0 & 0.0 & 0.0 & 0.0\\
  SPRM &0.6 &0.7 &0.5 &0.7 & 0.7 & 1.1 & 0.6 & 0.9\\
  
  \hline 
  
& \multicolumn{2}{c}{\textbf{E}} & \multicolumn{2}{c}{\textbf{F}} & \multicolumn{2}{c}{\textbf{G}} & \multicolumn{2}{c}{\textbf{H}}\\
 \cmidrule(lr{1em}){2-3} \cmidrule(lr{1em}){4-5} \cmidrule(lr{1em}){6-7} \cmidrule(lr{1em}){8-9}  
   Method& mean & sd & mean & sd & mean & sd & mean & sd\\
 \midrule
  {DCOV-SVS}  &7977.4 & 5911.1   & 6271.1&4782.4  &8758.4 &6821.6 &18325.7 &8778.8  \\
  MDD-SVS &11362.3& 7813.9 & \textbf{27956.3} &16236.9 &11904.2 &7859.0 &\textbf{29137.4} &{15116.9}  \\
  BCOV-SVS &\textbf{19217.2} &7952.6 &10768.4 &2271.3 &\textbf{20393.8} & 9617.5 &13511.9 &4319.6 \\
  SNIPLS &0.1 &0.0 &0.0 & 0.0 &0.0 &0.0 &0.0 &0.0\\
  SPRM &0.6 &0.9 &0.6 & 0.9 &0.6 & 0.8 &0.5 & 0.7
\\
  \hline\hline
\end{tabular}
\caption{Mean and standard deviation of time taken(s). \label{table:svstime} }
\end{sidewaystable}

\section{Real data application}
\label{sec:Example}

\subsection{Boston housing data}

The Boston housing data \citep{HARRISON1978} present a well studied data set in regression and dimension reduction analysis, which allows to compare the results to prior literature. Many reports have been published on the data, but generally an objective of interest when analyzing these data, is to understand how median house prices are impacted by a set of predictors in each of the 506 census tracts that the Boston Standard Metropolitan Statistical Area was composed of. This list of properties describing each census tract, is listed in Table \ref{tab:BostonHousingPredictors}. 

The Boston Housing data are known to be skewed, particularly with respect to crime rate, which can be mitigated by removing observations greater than 3.2 for that variable, as was also suggested in \cite{ChenShenYinESSDR2018}. Upon trimming, 374 observations out of the original 506 census tracts, remain. 
\begin{table}[!ht]
\centering
\begin{tabular}{cc}
\hline
Variable \# & Neighbourhood property \\
\hline
1 & per capita crime rate by town\\ 
2 & proportion of residential land zoned for lots over 25,000 sf\\ 
3 & proportion of nonretail business acres per town\\ 
4 & Charles River proximity (dichotomous)\\
5 & nitric oxide concentration in ambient air\\
6 & average number of rooms per dwelling\\
7 & proportion of owner-occupied units built prior to 1940\\ 
8 & weighted distances to five Boston employment centers\\ 
9 & index of accessibility to radial highways\\  
10 & full-value property-tax rate\\ 
11 & pupil-teacher ratio by town\\ 
12 & proportion of African Americans per town\\ 
13 & percentage of lower status of the population\\
\hline 
\end{tabular}
\caption{\label{tab:BostonHousingPredictors} Description of predictors in the Boston Housing data set}
\end{table}

To understand how the variables listed in Table \ref{tab:BostonHousingPredictors} bear relation to the median house price per census tract, the central subspace can be estimated, which by definition \eqref{eq:SDR} contains all information relevant to describe the dependent variable. However, in this analysis, one cannot rule out that not all of the predictors originally considered, should contribute the central subspace, as was surmised by \cite{ChenShenYinESSDR2018} as well. These authors proceeded to apply sparse DCOV---SDR in order to obtain a sparse estimate of the central subspace and report the resulting coefficients. These results will now be compared to sparse MDD---SDR and sparse BCOV---SDR.   

Following \cite{zhou2008}, the dimension of the central subspace of the Boston Housing data is $h=2$.  The coefficients of the loading vectors of the SDR central subspace estimates, are presented in table \ref{table:boston}. The optimal penalty parameters for DCOV---SVS, MDD---SVS and BCOV---SVS, were estimated to be $0.49$, $0.48$ and $0.17$, respectively.

\begin{table}[!htbp]
\centering
\begin{tabular}{ccccccc}
\hline\hline
Method & \multicolumn{2}{c}{\textbf{DCOV-SVS}} & \multicolumn{2}{c}{\textbf{MDD-SVS}} & \multicolumn{2}{c}{\textbf{BCOV-SVS}}\\
\hline
$x_1$  &0&0  &0&0 &0&0 \\
$x_2$  &0.268 &0.040 &0.310 &0.198 &0.006 &-0.0002 \\
$x_3$ & 0 &0  &-0.044 &-0.028 &0 &0 \\
$x_4$ & 0&0 &0&0 &0&0 \\
$x_5$ & 0&0 &0&0 &0&0 \\
$x_6$ &0&0 &0&0 &0&0 \\
$x_7$ &-0.240 &-0.037 &-0.246 &-0.157 &-0.001&0.001 \\
$x_8$ &0 &0 &0&0 &0&0 \\
$x_9$ &0 &0 &0&0 &0&0 \\
$x_{10}$ &-0.874 &-0.131 & -0.687 &-0.438 &0.001&0.001\\
$x_{11}$ &0 &0 &-0.017 &-0.011 &0&0 \\
$x_{12}$ & 0.159 &0.024 &0.179 &0.114 &-0.392 &0.047 \\
$x_{13}$ & -0.162 &-0.024 &-0.175 &-0.112 &0.132 &0.572\\
  \hline\hline
\end{tabular}
\caption{Estimated  SVS basis of the central subspace. \label{table:boston} }
\end{table}

Due to the trimming of the crime rate variable, it makes sense that that variable no longer significantly contributes to the central subspace. Furthermore, further variables that do not contribute to explaining the median house prices are: Charles river proximity, nitric oxide concentration in ambient air, the average number of rooms per dwelling, weighted distances to employment centers, as well as radial highways. Up to some extent, one can interpret why these variables could be less important than the other ones set forward. The BCOV and DCOV based variants add two additional variables to the list: the proportion of nonretail business area and the pupil---teacher ratio, whereas the MDD based variant does capture these, which illustrates that conditional mean independence does not have the same practical implications as distributional independence. MDD does add both variables with a negative sign, which corresponds to what one would, maybe subjectively, expect. 
Finally, it is noted that BCOV estimates some variables with an opposite sign compared to both other techniques. These are the proportion of African Americans in the population, which as hard to interpret as this may be in 2020, may have been seen as a negative factor back in the seventies. Also, BCOV estimates the proportion of poorer residents to be a positive instead of a negative. It is hard to say what this should be on an average basis. A high ratio of low status residents may indicate disrepair, but on a local level it may correspond to an "up-and-coming" redeveloping neighbourhood as well. The BCOV measure of association being a more local estimate, based on a $n$-ball around each point, can yield a different overall outcome.

\subsection{Riboflavin data}

\begin{figure}
\centering
\includegraphics[width=0.7\textwidth]{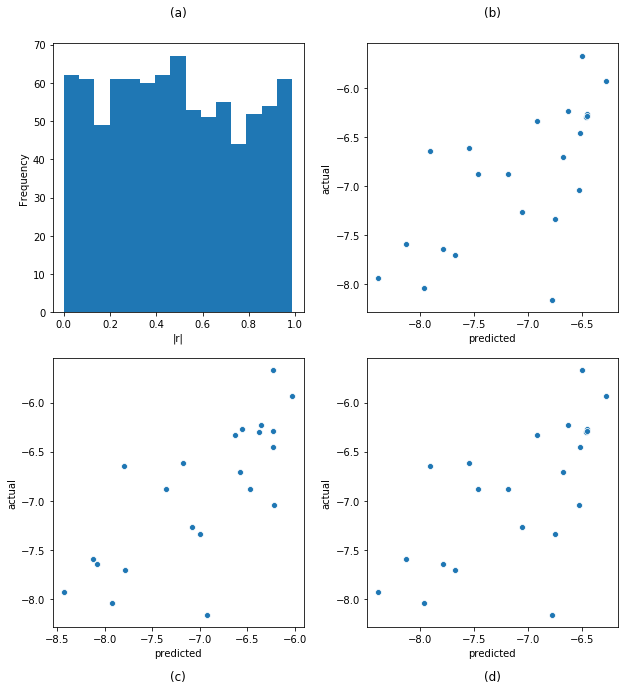}
\caption{Panel (a) is the histogram of the absolute correlations for the 854 selected genes; Panel (b) shows the test set predicted response vs actual response for DCOV---SVS ;Panel (c) shows the test set predicted response vs actual response for MDD---SVS ; Panel (d) shows the test set predicted response vs actual response for BCOV---SVS.}
\label{fig:riboflavin}
\end{figure}

To illustrate a high dimensional setting ($n<p$), a genomics example will be put forward. The study, originally published by \cite{buhlmann2014}, aims to model riboflavin production as the
single real-valued response variable from 4088 variables that
measure the logarithm of the expression level of the corresponding genes. Roboflavin production is reported as the logarithm of the riboflavin production rate.  The data consist of $n = 71$ samples that were hybridized repeatedly during a fed-batch fermentation process, in which different engineered
strains and strains grown under different fermentation conditions were studied. 
The data were split randomly into a training set of $49$ samples and a test set of $22$ samples. Next, in order to reduce the dimensionality to tractable levels, the training set was subjected to marginal variable screening based on the distance correlation $t$-test of independence \citep{SZEKELY2013193} on a pairwise basis between the predictors and the response. Variables that yield a $p$-value lower than 0.01 were retained, effectively reducing the number of variables from 4088 to 854.  Sparse SDR was applied \ref{sparseSDRalgo} to the dimension reduced training set, and the central subspace dimension was set to $h = 1$, as previously reported by \cite{YinHilafu2017}. Figure \ref{fig:riboflavin}(a) illustrates that the absolute correlation between the predictors is high, which further complicates the task of estimating the sparse central subspace. Figures \ref{fig:riboflavin} (b), (c), (d) show the predicted and actual response for DCOV---SVS, MDD---SVS and BCOV---SVS, respectively.  The predicted responses are obtained by fitting a linear regression to the latent variables that span the central subspace. This approach is analogous to e.g. principal component regression and can be used to predict the responses in the test set. Table \ref{table:riboflavin} confirms that all sparse solutions produce a better fit to the data than the non sparse solutions and are more parsimonious, hence justifying the need for sparse variable selection.  MDD---SVS produces a better fit to the data than DCOV---SVS and BCOV---SVS. This is further illustrated in table \ref{table:riboflavin} below, where MDD---SVS has the lowest Median Absolute Error (MAE). Note that the same pattern was present in the simulations of section \ref{sec:sim}: BCOV---SVS selects the lowest penalty parameter and the lowest number of predictors. 
Again, conditional mean independence presents a dimension reduction strategy different from distributional independence. From this example, as well as from the simulation study, it seems that MDD---SDR tends to lead to a less sparse central subspace than BCOV---SDR. In bioinformatics, a phenomenon like riboflavin production is rarely caused by the expression of a single gene, which in this case, seems to advocate for the MDD based approach from a practical perspective.

\begin{table}
\centering
\begin{tabular}{cccc}
\hline\hline
Method & \multicolumn{1}{c}{\textbf{DCOV---SVS}} & \multicolumn{1}{c}{\textbf{MDD---SVS}} & \multicolumn{1}{c}{\textbf{BCOV---SVS}}\\
\hline
Sparse MAE & 0.376 & 0.316 & 0.532\\
Non Sparse MAE & 0.378 &0.447 &0.666\\
$\theta$ & 0.22 & 0.10 & 0.01\\
Number of predictors & 3 &17 &1 \\
  \hline\hline
\end{tabular}
\caption{Median Absolute Error(MAE), penalty parameter $\theta$ and number of selected variables for  SVS methods. \label{table:riboflavin} }
\end{table}

The 17 predictors selected by MDD-SVS are: 'ASD\_at', 'DACA\_at', 'ECSC\_at', 'FLIJ\_at', 'FLIY\_at', 'FLIZ\_at', 'FRUB\_at', 'FTSE\_at', 'GSIB\_at', 'GUAB\_at', 'MURF\_at', 'MUTL\_at', 'PKSC\_at', 'PKSD\_at', 'PKSF\_at', 'PKSG\_at', 'PKSH\_at'. The variables selected by the two other methods are 'GUAB\_at', 'HEMA\_at', 'PKSD\_at' for DCOV-SVS and 'MRAY\_at' for BCOV-SVS.

\section{Conclusion and Outlook}
\label{sec:conc}
In this article, sparse sufficient dimension reduction methods have been presented based on energy and ball statistics. The advantage of these methods is that they are essentially nonparametric: they can be applied with little distributional assumptions. Three approaches were investigated: sparse sufficient dimension reduction that maximizes distance covariance, martingale difference divergence, or ball covariance. Whereas the first of these approaches had been reported before \citep{ChenShenYinESSDR2018}, the two remaining approaches have been put forward in this article. The paper has presented an extensive investigation of how these approaches compare, both in a simulation setting that compares different scenarios: linear and nonlinear dependencies, with clean data and data that contain outliers, as well as in two real data examples. The approaches are based on different statistical assumptions, each of which have their pros and cons. Therefore, one cannot expect a clear winner from these investigations. However, some take home messages are that ball covariance tends to lead to a more sparse estimate of the central subspace and may account for local variations in the data. On the contrary, zero martingale difference divergence guarantees conditional mean independence, which is an overall location property, and therefore looks at the data as a whole. MDD tends to report less sparse solutions, which can be an advantage when it makes sense to interpret an interplay of variables, such as in the riboflavin production example shown. 
However, some statement can be made across the different methods. All SDR methods investigated have been shown to be able to detect nonlinear dependencies. The DCOV and BCOV based options proved to lead to more accurate SVS in highly nonlinear contexts. They do work with little distributional information, yet are still sensitive to outliers. To their advantage, each of the methods can be applied equally well to data with more variables than predictors and the other way around. Howbeit, while the implementations described here and in \cite{ChenShenYinESSDR2018} may be the most efficient algorithms available, all three methods tend to be computationally involved. Therefore, we see the methods find immediate application in data sets of moderate sizes. Yet further research may bring an even more effient algorithm, such that the methods may find more widespread adoption. Future research could focus as well on more robust optimization algorithms for solving problem (\ref{eq:svsass_empir}) and different statistics that could be used in Algorithm \ref{sparseSDRalgo}.  


\bibliography{bibS27}

\begin{thebibliography}{36}
\expandafter\ifx\csname natexlab\endcsname\relax\def\natexlab#1{#1}\fi
\providecommand{\url}[1]{\texttt{#1}}
\providecommand{\href}[2]{#2}
\providecommand{\path}[1]{#1}
\providecommand{\DOIprefix}{doi:}
\providecommand{\ArXivprefix}{arXiv:}
\providecommand{\URLprefix}{URL: }
\providecommand{\Pubmedprefix}{pmid:}
\providecommand{\doi}[1]{\href{http://dx.doi.org/#1}{\path{#1}}}
\providecommand{\Pubmed}[1]{\href{pmid:#1}{\path{#1}}}
\providecommand{\bibinfo}[2]{#2}
\ifx\xfnm\relax \def\xfnm[#1]{\unskip,\space#1}\fi
\bibitem[{Bühlmann et~al.(2014)Bühlmann, Kalisch and Meier}]{buhlmann2014}
\bibinfo{author}{Bühlmann, P.}, \bibinfo{author}{Kalisch, M.},
  \bibinfo{author}{Meier, L.}, \bibinfo{year}{2014}.
\newblock \bibinfo{title}{High-dimensional statistics with a view toward
  applications in biology}.
\newblock \bibinfo{journal}{Annual Review of Statistics and Its Application}
  \bibinfo{volume}{1}, \bibinfo{pages}{255--278}.
\bibitem[{Chen et~al.(2018)Chen, Shen and Yin}]{ChenShenYinESSDR2018}
\bibinfo{author}{Chen, X.}, \bibinfo{author}{Shen, W.}, \bibinfo{author}{Yin,
  X.}, \bibinfo{year}{2018}.
\newblock \bibinfo{title}{Efficient sparse estimate of sufficient dimension
  reduction in high dimension}.
\newblock \bibinfo{journal}{Technometrics} \bibinfo{volume}{60},
  \bibinfo{pages}{161--168}.
\bibitem[{Chen et~al.(2010)Chen, Zou and Cook}]{chen2010}
\bibinfo{author}{Chen, X.}, \bibinfo{author}{Zou, C.}, \bibinfo{author}{Cook,
  R.}, \bibinfo{year}{2010}.
\newblock \bibinfo{title}{Coordinate-independent sparse sufficient dimension
  reduction and variable selection}.
\newblock \bibinfo{journal}{The Annals of Statistics} \bibinfo{volume}{38},
  \bibinfo{pages}{3696--3723}.
\bibitem[{Conover(1999)}]{Conover1999}
\bibinfo{author}{Conover, W.}, \bibinfo{year}{1999}.
\newblock \bibinfo{title}{Practical nonparametric statistics}.
\newblock Wiley series in probability and statistics. \bibinfo{edition}{3. ed}
  ed., \bibinfo{publisher}{Wiley}, \bibinfo{address}{New York, NY [u.a.]}.
\bibitem[{Cook(2004)}]{Cook2004}
\bibinfo{author}{Cook, R.}, \bibinfo{year}{2004}.
\newblock \bibinfo{title}{Testing predictor contributions in sufficient
  dimension reduction}.
\newblock \bibinfo{journal}{The Annals of Statistics} \bibinfo{volume}{32},
  \bibinfo{pages}{1062--1092}.
\bibitem[{Cook et~al.(2016)Cook, Helland and Su}]{CookHellandSu2013}
\bibinfo{author}{Cook, R.}, \bibinfo{author}{Helland, I.}, \bibinfo{author}{Su,
  Z.}, \bibinfo{year}{2016}.
\newblock \bibinfo{title}{Scaled predictor envelopes and partial least-squares
  regression}.
\newblock \bibinfo{journal}{Technometrics} \bibinfo{volume}{11},
  \bibinfo{pages}{155--165}.
\bibitem[{Fan and Li(2001)}]{FanLi2001}
\bibinfo{author}{Fan, J.}, \bibinfo{author}{Li, R.}, \bibinfo{year}{2001}.
\newblock \bibinfo{title}{Variable selection via nonconcave penalized
  likelihood and its oracle properties}.
\newblock \bibinfo{journal}{Journal of the American Statistical Association}
  \bibinfo{volume}{96}, \bibinfo{pages}{1348--1360}.
\bibitem[{Fan and Lv(2008)}]{FanLvSIS2008}
\bibinfo{author}{Fan, J.}, \bibinfo{author}{Lv, J.}, \bibinfo{year}{2008}.
\newblock \bibinfo{title}{Sure independence screening for ultrahigh dimensional
  feature space}.
\newblock \bibinfo{journal}{Journal of the Royal Statistical Society, Series B}
  \bibinfo{volume}{70}, \bibinfo{pages}{849--911}.
\bibitem[{Harrison and Rubinfeld(1978)}]{HARRISON1978}
\bibinfo{author}{Harrison, D.}, \bibinfo{author}{Rubinfeld, D.L.},
  \bibinfo{year}{1978}.
\newblock \bibinfo{title}{Hedonic housing prices and the demand for clean air}.
\newblock \bibinfo{journal}{Journal of Environmental Economics and Management}
  \bibinfo{volume}{5}, \bibinfo{pages}{81 -- 102}.
\bibitem[{Hilafu and Yin(2017)}]{YinHilafu2017}
\bibinfo{author}{Hilafu, H.}, \bibinfo{author}{Yin, X.}, \bibinfo{year}{2017}.
\newblock \bibinfo{title}{Sufficient dimension reduction and variable selection
  for large-p-small-n data with highly correlated predictors}.
\newblock \bibinfo{journal}{Journal of Computational and Graphical Statistics}
  \bibinfo{volume}{26}, \bibinfo{pages}{26--34}.
\bibitem[{Hoffmann et~al.(2016)Hoffmann, Filzmoser, Serneels and
  Varmuza}]{SPRM-DA}
\bibinfo{author}{Hoffmann, I.}, \bibinfo{author}{Filzmoser, P.},
  \bibinfo{author}{Serneels, S.}, \bibinfo{author}{Varmuza, K.},
  \bibinfo{year}{2016}.
\newblock \bibinfo{title}{Sparse and robust {PLS} for binary classification}.
\newblock \bibinfo{journal}{Journal of Chemometrics} \bibinfo{volume}{30},
  \bibinfo{pages}{153--162}.
\bibitem[{Hoffmann et~al.(2015)Hoffmann, Serneels, Filzmoser and Croux}]{SPRM}
\bibinfo{author}{Hoffmann, I.}, \bibinfo{author}{Serneels, S.},
  \bibinfo{author}{Filzmoser, P.}, \bibinfo{author}{Croux, C.},
  \bibinfo{year}{2015}.
\newblock \bibinfo{title}{Sparse partial robust {M} regression}.
\newblock \bibinfo{journal}{Chemometrics and Intelligent Laboratory Systems}
  \bibinfo{volume}{149}, \bibinfo{pages}{50--59}.
\bibitem[{Huber(1985)}]{Huber1985}
\bibinfo{author}{Huber, P.J.}, \bibinfo{year}{1985}.
\newblock \bibinfo{title}{Projection pursuit}.
\newblock \bibinfo{journal}{The Annals of Statistics} \bibinfo{volume}{13},
  \bibinfo{pages}{435--475}.
\bibitem[{Huo and Sz\'{e}kely(2016)}]{HuoFastDCOV2016}
\bibinfo{author}{Huo, X.}, \bibinfo{author}{Sz\'{e}kely, G.},
  \bibinfo{year}{2016}.
\newblock \bibinfo{title}{Fast computing for distance covariance}.
\newblock \bibinfo{journal}{Technometrics} \bibinfo{volume}{58},
  \bibinfo{pages}{435--447}.
\bibitem[{Knyazev and Argentati(2002)}]{Knyazev2002PrincipalAB}
\bibinfo{author}{Knyazev, A.}, \bibinfo{author}{Argentati, M.},
  \bibinfo{year}{2002}.
\newblock \bibinfo{title}{Principal angles between subspaces in an a-based
  scalar product: Algorithms and perturbation estimates}.
\newblock \bibinfo{journal}{SIAM J. Sci. Comput.} \bibinfo{volume}{23},
  \bibinfo{pages}{2008--2040}.
\bibitem[{Kruskal and Wallis(1952)}]{KruskalWallis1952}
\bibinfo{author}{Kruskal, W.H.}, \bibinfo{author}{Wallis, W.A.},
  \bibinfo{year}{1952}.
\newblock \bibinfo{title}{Use of ranks in one-criterion variance analysis}.
\newblock \bibinfo{journal}{Journal of the American Statistical Association}
  \bibinfo{volume}{47}, \bibinfo{pages}{583--621}.
\bibitem[{Li(2018)}]{LiSDR}
\bibinfo{author}{Li, B.}, \bibinfo{year}{2018}.
\newblock \bibinfo{title}{Sufficient Dimension Reduction: Methods and
  Applications with R}.
\newblock \bibinfo{publisher}{Chapman {\&} Hall /CRC, Monographs on Statistics
  and Applied Probability}, \bibinfo{address}{New York}.
\bibitem[{Li(2007)}]{LiSVS2007}
\bibinfo{author}{Li, L.}, \bibinfo{year}{2007}.
\newblock \bibinfo{title}{Sparse sufficient dimension reduction}.
\newblock \bibinfo{journal}{Biometrika} \bibinfo{volume}{94},
  \bibinfo{pages}{603--613}.
\bibitem[{Lyons(2013)}]{Lyons2013}
\bibinfo{author}{Lyons, R.}, \bibinfo{year}{2013}.
\newblock \bibinfo{title}{Distance covariance in metric spaces}.
\newblock \bibinfo{journal}{Annals of Probability} \bibinfo{volume}{41},
  \bibinfo{pages}{3284--3305}.
\bibitem[{{Menvouta} et~al.(2020){Menvouta}, {Serneels} and
  {Verdonck}}]{menvouta2020}
\bibinfo{author}{{Menvouta}, E.J.}, \bibinfo{author}{{Serneels}, S.},
  \bibinfo{author}{{Verdonck}, T.}, \bibinfo{year}{2020}.
\newblock \bibinfo{title}{{direpack: A Python 3 package for state-of-the-art
  statistical dimension reduction methods}}.
\newblock \bibinfo{journal}{arXiv e-prints} ,
  \bibinfo{pages}{arXiv:2006.01635}.
\bibitem[{Pan et~al.(2019)Pan, Wang, Xiao and Zhu}]{PanWXZ2019}
\bibinfo{author}{Pan, W.}, \bibinfo{author}{Wang, X.}, \bibinfo{author}{Xiao,
  W.}, \bibinfo{author}{Zhu, H.}, \bibinfo{year}{2019}.
\newblock \bibinfo{title}{A generic sure independence screening procedure}.
\newblock \bibinfo{journal}{Journal of the American Statistical Association}
  \bibinfo{volume}{114}, \bibinfo{pages}{928--937}.
\bibitem[{Pan et~al.(2020)Pan, Wang, Zhang, Zhu and Zhu}]{PanWXZ2020BCov}
\bibinfo{author}{Pan, W.}, \bibinfo{author}{Wang, X.}, \bibinfo{author}{Zhang,
  H.}, \bibinfo{author}{Zhu, H.}, \bibinfo{author}{Zhu, J.},
  \bibinfo{year}{2020}.
\newblock \bibinfo{title}{Ball covariance: A generic measure of dependence in
  banach space}.
\newblock \bibinfo{journal}{Journal of the American Statistical Association}
  \bibinfo{volume}{115}, \bibinfo{pages}{307--317}.
\bibitem[{Park et~al.(2015)Park, Shao and Yao}]{park2015}
\bibinfo{author}{Park, T.}, \bibinfo{author}{Shao, X.}, \bibinfo{author}{Yao,
  S.}, \bibinfo{year}{2015}.
\newblock \bibinfo{title}{Partial martingale difference correlation}.
\newblock \bibinfo{journal}{Electronic Journal of Statistics}
  \bibinfo{volume}{9}, \bibinfo{pages}{1492--1517}.
\bibitem[{Raymaekers and Rousseeuw(2019)}]{RaymaekersRousseeuw2019}
\bibinfo{author}{Raymaekers, J.}, \bibinfo{author}{Rousseeuw, P.J.},
  \bibinfo{year}{2019}.
\newblock \bibinfo{title}{A generalized spatial sign covariance matrix}.
\newblock \bibinfo{journal}{Journal of Multivariate Analysis}
  \bibinfo{volume}{171}, \bibinfo{pages}{94--111}.
\bibitem[{Serneels et~al.(2005)Serneels, Croux, Filzmoser and Van~Espen}]{PRM}
\bibinfo{author}{Serneels, S.}, \bibinfo{author}{Croux, C.},
  \bibinfo{author}{Filzmoser, P.}, \bibinfo{author}{Van~Espen, P.J.},
  \bibinfo{year}{2005}.
\newblock \bibinfo{title}{Partial robust {M}-regression}.
\newblock \bibinfo{journal}{Chemometrics and Intelligent Laboratory Systems}
  \bibinfo{volume}{79}, \bibinfo{pages}{55--64}.
\bibitem[{Shao and Zhang(2014)}]{ShaoYangMDD2014}
\bibinfo{author}{Shao, X.}, \bibinfo{author}{Zhang, J.}, \bibinfo{year}{2014}.
\newblock \bibinfo{title}{Martingale difference correlation and its use in
  high-dimensional variable screening}.
\newblock \bibinfo{journal}{Journal of the American Statistical Association}
  \bibinfo{volume}{109}, \bibinfo{pages}{1302--1318}.
\bibitem[{Sheng and Yin(2016)}]{ShengYinDCOVSDR2016}
\bibinfo{author}{Sheng, W.}, \bibinfo{author}{Yin, X.}, \bibinfo{year}{2016}.
\newblock \bibinfo{title}{Sufficient dimension reduction via distance
  covariance}.
\newblock \bibinfo{journal}{Journal of Computational and Graphical Statistics}
  \bibinfo{volume}{25}, \bibinfo{pages}{91--104}.
\bibitem[{Sz\'{e}kely and Rizzo(2013)}]{SzekelyRizzo2013}
\bibinfo{author}{Sz\'{e}kely, G.}, \bibinfo{author}{Rizzo, M.},
  \bibinfo{year}{2013}.
\newblock \bibinfo{title}{Energy statistics: A class of statistics based on
  distances}.
\newblock \bibinfo{journal}{Journal of Statistical Planning and Inference}
  \bibinfo{volume}{143}, \bibinfo{pages}{1249--1272}.
\bibitem[{Sz\'{e}kely et~al.(2007)Sz\'{e}kely, Rizzo and
  Bakirov}]{SzekelyRizzoBakirov2007}
\bibinfo{author}{Sz\'{e}kely, G.}, \bibinfo{author}{Rizzo, M.},
  \bibinfo{author}{Bakirov, N.}, \bibinfo{year}{2007}.
\newblock \bibinfo{title}{Measuring and testing dependence by correlation of
  distances}.
\newblock \bibinfo{journal}{The Annals of Statistics} \bibinfo{volume}{35},
  \bibinfo{pages}{2769--2794}.
\bibitem[{Székely and Rizzo(2013)}]{SZEKELY2013193}
\bibinfo{author}{Székely, G.J.}, \bibinfo{author}{Rizzo, M.L.},
  \bibinfo{year}{2013}.
\newblock \bibinfo{title}{The distance correlation t-test of independence in
  high dimension}.
\newblock \bibinfo{journal}{Journal of Multivariate Analysis}
  \bibinfo{volume}{117}, \bibinfo{pages}{193 -- 213}.
\bibitem[{W\"{a}chter and Biegler(2006)}]{WaechterBiegler2006}
\bibinfo{author}{W\"{a}chter, A.}, \bibinfo{author}{Biegler, L.T.},
  \bibinfo{year}{2006}.
\newblock \bibinfo{title}{On the implementation of an interior-point filter
  line-search algorithm for large-scale nonlinear programming}.
\newblock \bibinfo{journal}{Mathematical Programming} \bibinfo{volume}{106},
  \bibinfo{pages}{25--57}.
\bibitem[{Yin and Hilafu(2015)}]{YinHilafu2015}
\bibinfo{author}{Yin, X.}, \bibinfo{author}{Hilafu, H.}, \bibinfo{year}{2015}.
\newblock \bibinfo{title}{Sequential sufficient dimension reduction for large
  p, small n problems}.
\newblock \bibinfo{journal}{Journal of the Royal Statistical Society. Series B
  (Statistical Methodology)} \bibinfo{volume}{77}, \bibinfo{pages}{879--892}.
\bibitem[{Zhang and Chen(2019)}]{ZhangChenBCOVSDR2019}
\bibinfo{author}{Zhang, J.}, \bibinfo{author}{Chen, X.}, \bibinfo{year}{2019}.
\newblock \bibinfo{title}{Robust sufficient dimension reduction via ball
  covariance}.
\newblock \bibinfo{journal}{Computational Statistics and Data Analysis}
  \bibinfo{volume}{140}, \bibinfo{pages}{144--154}.
\bibitem[{Zhang et~al.(2019)Zhang, Liu, Wu and Fang}]{ZhangMDDSDR2019}
\bibinfo{author}{Zhang, Y.}, \bibinfo{author}{Liu, J.}, \bibinfo{author}{Wu,
  Y.}, \bibinfo{author}{Fang, X.}, \bibinfo{year}{2019}.
\newblock \bibinfo{title}{A martingale-difference-divergence-based estimation
  of central mean subspace}.
\newblock \bibinfo{journal}{Statistics and Its Interface} \bibinfo{volume}{12},
  \bibinfo{pages}{489--500}.
\bibitem[{Zhou and He(2008)}]{zhou2008}
\bibinfo{author}{Zhou, J.}, \bibinfo{author}{He, X.}, \bibinfo{year}{2008}.
\newblock \bibinfo{title}{Dimension reduction based on constrained canonical
  correlation and variable filtering}.
\newblock \bibinfo{journal}{Ann. Statist.} \bibinfo{volume}{36},
  \bibinfo{pages}{1649--1668}.
\bibitem[{Zou et~al.(2006)Zou, Hastie and Tibshirani}]{ZouSPCA2006}
\bibinfo{author}{Zou, H.}, \bibinfo{author}{Hastie, T.},
  \bibinfo{author}{Tibshirani, R.}, \bibinfo{year}{2006}.
\newblock \bibinfo{title}{Sparse principal component analysis}.
\newblock \bibinfo{journal}{Journal of Computational and Graphical Statistics}
  \bibinfo{volume}{15}, \bibinfo{pages}{265--286}.

\end{thebibliography}

\end{document}